\documentclass[prb,aps,twocolumn]{revtex4}

\usepackage{dcolumn}
\usepackage{bm}
\usepackage{times}
\usepackage[]{graphicx}

\begin{document}




%
\newcommand{\rys}[3]{\begin{figure}
\includegraphics[width=.49\textwidth]{#1}
\caption{\label{#2} #3}
\end{figure}}

\title{ Magnetic flux in mesoscopic rings: Quantum Smoluchowski regime
}
\author{  J. Dajka,  L. Machura,  Sz. Rogozi\'{n}ski and 
J. \L uczka }
 \affiliation{Institute of Physics, University of Silesia, 
    40-007 Katowice, Poland }
\date{\today}
\begin{abstract} 
Magnetic flux in mesoscopic rings under the quantum Smoluchowski regime is investigated. Quantum corrections to the dissipative current are shown to form multistable steady states and can result in statistical enhancement of the magnetic flux.   The relevance of quantum correction  effects is supported  via the  entropic criterion. A possible application  for a qutrit architecture  of quantum information is proposed.
\end{abstract}

\maketitle

\section{Introduction}

Mesoscopic systems belong to one of the most intriguing part  of present-day 
investigations. They occupy territory   between physics of small quantum objects 
and physics of  macroscopic objects. Many aspects of that territory remains {\it terra incognita} both
to experimentalists and theoreticians. 
For example, over one decade after first experiments \cite{buh,maily,mohanty}  proving the 
existence of the theoretically predicted \cite{hund,but} persistent 
currents in normal metal multiply connected samples, there is an 
unsolved central question: which mechanism is responsible for the 
unexpectedly large amplitude of the measured  current \cite{szwab}.
There is a suggestion  that the large current is due to  
{\it non-equilibrium} noise presented in the system \cite{kraw}. 
It is also theoretically predicted  \cite{koh} that currents in mesoscopic rings can flow even in  absence of any driving. Such {\it self-sustaining} currents  has not been observed so far. Their existence is a desired property for the  quantum information retrieval and computing technologies based on non-superconducting devices \cite{fq}. 

In our earlier work \cite{daj1} we proposed the two-fluid 
model of noisy dynamics of the magnetic flux 
in  mesoscopic rings and  cylinders.   
Dynamics of the magnetic flux 
is described by an  evolution equation which is equivalent to a Langevin  equation for an overdamped motion of a classical  
Brownian particle and a steady state of the system is  characterized by 
the asymptotic probability density being a stationary solution of  the corresponding Fokker-Planck equation. In this approach,  self-sustaining fluxes are long living states of the system described by a multistable asymptotic probability density. This model is an example of a hybrid of quantum and classical parts and is a counterpart  of the well known model of a resistively shunted Josephson junction \cite{barone}. The classical part consists of 'normal' electrons carrying dissipative current. The  quantum part is formed by those electrons which maintain their phase coherence around the circumference of the cylinder or ring. The effective kinetics is determined by a  classical Langevin equation 
with a Nyquist noise describing thermal equilibrium fluctuations. The coherent part of the system acts as an  additional 'force' driving normal electrons.    It is natural to ask what is an impact of quantum nature of dissipative kinetics on the properties of fluxes and currents flowing in such systems. 
To answer this question, we exploit the  
so called Quantum Smoluchowski Equation  introduced in Ref. \onlinecite{ankerhold1} and, 
with the  Maxwell daemon successfully exorcised,   
 in Refs.\onlinecite{luczka,ankerhold2}.    
First, we  extend our model for  overdamped kinetics \cite{daj1} to the domain where charging effects (corresponding to the inertial effects for particles) appear. This extension is necessary for a precise  identification  of the quantum Smoluchowski regime.  The quantum corrections are of great importance for the existence and properties of self-sustaining currents or magnetic fluxes. It is shown below that in moderate, with respect to the gap at the Fermi level, temperatures these quantum corrections are destructive for their existence. It is not the case at lower temperature: one gets not only the multistability of the probability density but also significant enhancement of the probability of the occurrence of long living states carrying magnetic  flux of a certain amplitude. 

It is shown that for the system under consideration  the passage from the classical Smoluchowski regime  into the quantum Smoluchowski regime is accompanied with decrease of the Shannon entropy. It emphasizes the significance of the multistable ordered state. As the predicted multistability is formed by a set of odd number of maxima in the asymptotic probability density it is natural to expect the ring or cylinder to be a candidate for a {\it qutrit} rather than a qubit. 
 
The layout of the paper is  as follows: In Sec. II, we construct 
an extended, capacitive model of dissipative magnetic flux dynamics 
in mesoscopic systems of a cylinder symmetry. Next, in Sec. III, 
we discuss the quantum Smoluchowski regime for the system. In Sec. IV, we study properties of the stationary magnetic flux in the  
quantum Smoluchowski domain. Sec. V contains summary and conclusions.  


\section{Capacitive model of dissipative flux dynamics}  

At zero temperature $T$,
small metallic systems of the cylinder symmetry  (like rings, toroids and cylinders) threaded by a magnetic flux $\phi$ display 
persistent and non-dissipative  currents $I_{coh}$ run by  coherent
electrons. At  non-zero temperature,  a part of electrons
becomes 'normal' (non-coherent) and the amplitude of the
persistent current decreases. Moreover, 
 resistance of the ring and thermal fluctuations start to play a role.  Therefore for temperatures  $T>0$, there are both coherent  and dissipative 
parts  of the total current, namely,  
\begin{equation} \label{I}
I_{tot}=I_{coh} + I_{dis}.
\end{equation}
 The persistent 
current $I_{coh}$ as a function of the magnetic flux $\phi$ depends on the parity of
the number of coherent electrons. Let $p$ denotes the
probability of an even number of coherent electrons. Then the formula for coherent current  reads \cite{cheng}
\begin{eqnarray}
I_{coh}=I_{coh}(\phi,T)=p\,I_{even}(\phi,T)+(1-p)\,I_{odd}(\phi,T),
\end{eqnarray}
where
\begin{eqnarray}
I_{even}(\phi,T)=I_{odd}(\phi+\phi_0/2,T) =\nonumber\\
=I_0\sum_{n=1}^\infty A_n(T/T^*)\sin ( 2n\pi \phi /\phi _0).
\end{eqnarray}
 The flux quantum
$\phi_0=h/e$ is the ratio of the Planck constant $h$ and the charge of the electron,  $I_0$ is the maximal current at zero temperature.
 The temperature dependent amplitudes are determined by the relation 
\cite{cheng}
\begin{eqnarray}
A_n(T/T^*)= \frac{4T}{\pi T^*}\frac{\exp(-nT/T^*)}{1-\exp(-2nT/T^*)}
\cos(nk_F l),
\end{eqnarray}
 where the  characteristic temperature $T^*$ defined by the relation 
$k_BT^*=\Delta_F/2\pi^2$, where $\Delta _F$ is the energy gap 
at the Fermi surface, $k_B$ is the Boltzmann constant and  $k_F$ is the 
Fermi momentum and $l$ is the circumference of the ring. 

The dissipative current $I_{dis}$ is determined by
 the Ohm's law and Lenz's rule \cite{lucz}, 
\begin{equation}   \label{inor}
I_{dis} = I_{dis}(\phi, T)=-\frac{1}{R}\frac{d\phi}{dt}
+\sqrt{\frac{2k_BT}{R}}~\Gamma(t)\;,
\end{equation}
where $R$ is the resistance of the ring  and $\Gamma(t)$  models 
thermal Nyquist fluctuations of the Ohmic current. In the first approximation, this
thermal noise is  classical Gaussian white noise  of  zero average, i.e.,
$\langle \Gamma(t)\rangle=0$ and  $\delta$-auto-correlated function
$\langle\Gamma(t)\Gamma(s)\rangle=\delta(t-s)$. The
noise  intensity $D_0=\sqrt{2k_BT/R}$ is chosen in accordance with
the classical fluctuation-dissipation theorem.  

Quantum corrections to classical thermal fluctuations will be considered below in the so-called Smoluchowski regime. To define precisely this regime, first we have to include charging effects \cite{akty}. To this aim, we shall construct a formal  Hamilton function (i.e. energy)  of the system which 
 consists of three parts. The first one corresponds to an effective potential related to the persistent current itself;
 the second is related to the energy of the  magnetic flux and the third 
is due to  charging effects caused by capacitance $C$ of the system (it corresponds to the kinetic energy of a particle). 

We define a  potential energy  related to the persistent current  
 by the  relation 
\begin{equation}
E_{coh}(\phi)= - \int I_{coh}(\phi, T) d\phi,
\end{equation}
which reflects the well known fact that the persistent current is an 
equilibrium and thermodynamic phenomenon. 
At zero temperature, it is an energy of the set of discrete energy levels carrying persistent current.
For non-zero temperature, the persistent current is averaged over the  thermal distribution function and the above relation holds for a thermodynamic potential.

We assume that the ring can be characterized by a capacitance $C$. 
To justify it we cite 
Kopietz \cite{capac} who showed that in the diffusive regime the
 energy associated with long-wavelength 
and low-energy charge fluctuations is determined by classical charging 
energies and therefore the ring  behaves as it were a classical capacitor. 
 The flux dependence of these energies yields the contribution 
to the persistent current. 
The speculations that the local charge fluctuations and 
charging energies could contribute to persistent 
current has also been suggested  by Imry and Altshuler \cite{capac1}. 

From the above it follows that the total energy takes the form \cite{akty}
\begin{equation} \label{H}
E=\frac{C}{2}\left(\frac{d\phi}{dt}\right)^2+\frac{1}{2L}
\left(\phi-\phi_e\right)^2+E_{coh}(\phi),
\end{equation} 
where $\phi_e$ is the magnetic flux induced by an  external magnetic field $B$ and  $L$ is a self-inductance of the system. 
The equation of motion, which corresponds to (\ref{H}),  has the form
\begin{equation} \label{em}
C\frac{d^2\phi}{dt^2}=-\frac{1}{L}(\phi-\phi_e)+I_{coh}(\phi,T).
\end{equation}
Now, we want to take into account dissipation effects. To this aim we 
generalize Eq. (\ref{em}) replacing the coherent current $I_{coh}$ by the total current $I_{tot}$ given by (\ref{I}). 
As a result we obtain the evolution equation
\begin{eqnarray}\label{BE}
C\frac{d^2\phi}{dt^2} + \frac{1}{R}\frac{d\phi}{dt} &=& 
-\frac{1}{L}(\phi-\phi_{e}) + I_{coh}(\phi, T)
\nonumber\\ 
+ \sqrt{\frac{2 k_BT}{R}}\;\Gamma (t) 
 &=& - \frac{dW(\phi)}{d\phi} + \sqrt{\frac{2 k_BT}{R}}\;\Gamma (t),
\end{eqnarray}
where the  potential $W(\phi)$ reads 
\begin{eqnarray}\label{W}
W(\phi)&=&\frac{1}{2L} \left(\phi - \phi_e\right)^2 
\nonumber\\&+&
\phi_0 
I_0\sum_{n=1}^{\infty}
\frac{A_n(T/T^*)}{2n\pi} \left\{ 
p\cos\left(2n\pi \frac{\phi}{\phi_0}\right) \right. \nonumber\\
&+& \left. (1-p)\cos\left[2n\pi\left(
\frac{\phi}{\phi_0}+\frac{1}{2}\right)\right]\right\}.
\end{eqnarray}
This equation is  extended one in comparison with  the equation of motion 
studied in Ref. \onlinecite{daj1} by including the inertial, capacitive term. Its structure is similar to the model of capacitively and resistively shunted Josephson junction \cite{barone}. Indeed, the 
dynamics of a  trapped magnetic flux in a   superconducting ring interrupted by the Josephson junction is described by Eq. (\ref{BE}) by changing 
$I_{coh}(\phi, T)$ into the Josephson supercurrent $I=I_c \sin(\phi)$ 
\cite{weiss}.


\section{Quantum Smoluchowski regime}

The dissipative part of the current, given by Eq. (\ref{inor}), 
is classical one in which a quantum character of thermal fluctuations is ignored. 
At lower temperatures, it can be insufficient and leading quantum corrections might be important. We do not know how to incorporate quantum corrections in a general case described by (\ref{BE}).  
However, in the regimes where the charging effects can be neglected,  the system can be described by the so called quantum Smoluchowski equation 
\cite{ankerhold1,luczka}.  It has the same structure as a classical Smoluchowski equation, in which the potential $W(\phi))$ and diffusion coefficient $D_0=k_BT/R$ are modified due to quantum effects like tunneling, quantum reflections and fluctuations.  In terms of  the Langevin 
equation (\ref{BE}), it assumes the form
\begin{equation}\label{QOV}
\frac{1}{R}\frac{d\phi}{dt} = - \frac{dW_{m}(\phi)}{d\phi} +
\sqrt{2D_{m}(\phi)}\;\Gamma (t). 
\end{equation}
This equation has to be interpreted  in the Ito sense \cite{gard}.  
The  modified potential $W_{m}(\phi)$ and the modified
 diffusion coefficient $D_{m}(\phi)$ take the form \cite{luczka} 
\begin{eqnarray}
W_{m}(\phi)=W(\phi)+\frac{1}{2} \Lambda W''(\phi),\\
D_{m}(\phi)=\frac{D_0}{1-\Lambda W''(\phi)/k_BT}, 
\end{eqnarray}
where the prime denotes differentiation with respect to the argument of the function.  
The quantum corrections are characterized by  the parameter 
$\Lambda$. It measures a deviation of the quantal flux fluctuations from 
its classical counterpart, namely, 
\begin{equation}\label{Lam1}
\Lambda=\langle \phi^2\rangle_{Q} -\langle \phi^2\rangle_{C},
\end{equation}
where $\langle \cdots \rangle$ denotes equilibrium average, the subscripts 
$Q$ and $C$ refer to quantal and classical cases, respectively. The explicit 
form of $\Lambda$ reads \cite{talk} 
\begin{equation}\label{Lam2}
\Lambda= \frac{ \hbar R}{\pi}\left[ \Psi(1+\lambda_1/\nu) - 
\Psi(1+\lambda_2/\nu) \right],
\end{equation}
where the psi function $\Psi(z)$ is the logarithmic derivative of the 
Gamma function and 
\begin{eqnarray} \label{lam12}
\lambda_{1/2} = \omega_0\left[k\pm \sqrt{k^2 - 1} \right], \nonumber\\
k=(2\omega_0 CR)^{-1}, \quad \nu = 2\pi k_BT/\hbar.
\end{eqnarray}
The frequency $\omega_0$ is a typical frequency of the bare system 
and its inverse corresponds to a characteristic time of the system.  

Now, let us determine the range of applicability of the quantum Smoluchowski regime. The classical Smoluchowski limit corresponds to the neglect of charging effects. Formally,  
we should put $C=0$  in the inertial term of Eq. (\ref{BE}), which is 
related to the strong damping limit of the Brownian particle. In the
 case studied here it means that 
\begin{equation}\label{ineq1}
k \gg 1 \quad \mbox{or} \quad \omega_0 CR \ll 1
\end{equation}
and then   Eq. (\ref{Lam2}) takes the form 
\begin{equation}\label{Lam3}
\Lambda = \frac{ \hbar R}{\pi}\left[ \gamma + 
\Psi\left(1+\frac{\hbar}{2\pi CR k_BT}\right)  \right],
\end{equation}
where  $\gamma \simeq 0.5772$ is the Euler constant. 

The separation of time scales, on which the flux relaxes and the conjugate observable (a charge) \cite{luis} is already equilibrated, requires the second  condition, namely,  
\begin{equation}\label{ineq2}
\omega_0 CR \ll k_BT/\hbar \omega_0.
\end{equation}
 In the deep quantum regime, i.e. when 
\begin{equation}\label{ineq3}
 k_BT \ll \frac{\hbar}{2\pi CR},
\end{equation}
 the correction (\ref{Lam3}) assumes the form 
\begin{equation}\label{ln}
\Lambda=\frac{\hbar R}{\pi}\left[\gamma+\ln\left(\frac{\hbar}{2\pi CR k_BT}\right)\right]. 
\end{equation}
In order to identify precisely the  quantum Smoluchowski regime, we have to determine a typical frequency $\omega_0$ or the corresponding characteristic time $\tau_0 \propto 1/\omega_0$. There are many characteristic 
times in the system, which can be explicitly extracted from the 
evolution equation (\ref{BE}), e.g.  $CR, \hbar/ k_BT,  \phi_0/(RI_0)$. The characteristic time 
$\tau_0 = L/R$ is the relaxation time of the flux 
in the classical (non-coherent)  systems and below we scale time with  respect to $\tau_0$.  Why time is scaled in this way, 
we refer the readers to our previous paper \cite{akty}.  
Therefore, in the quantum Smoluchowski regime, 
all the above inequalities 
(\ref{ineq1}), (\ref{ineq2}) and (\ref{ineq3}) should be fulfilled 
for $\omega_0 \propto 1/\tau_0$. 
Because the diffusion coefficient cannot be negative,    the parameter $\Lambda$ should be chosen  small enough to satisfy 
the condition $D_{m}(\phi)\ge 0$ for all values of $\phi$.  
We note that the  passage from the classical Smoluchowski domain  to the quantum Smoluchowski domain allows for  the identification of the physical regime because of the formal similarities of the inertial and capacitive  terms in the equations of motion for the Brownian particle and the magnetic flux, respectively.  


\section{Steady state analysis}

From the mathematical point of view, 
the Langevin equation (\ref{QOV}) describes a classical 
Markov stochastic process. Therefore its all statistical properties can be obtained from the corresponding Fokker-Planck equation for the probability density. To analyze its stationary solution, 
let us introduce dimensionless variables in Eq. (\ref{QOV}):  
 the rescaled flux $x=\phi/\phi_0$ and rescaled  time $s =t/\tau_0$, 
where the characteristic time $\tau_0= L/R$. 
Then Eq. (11) can be rewritten in the dimensionless form  
\begin{equation}\label{LL}
\frac{dx}{ds}
=-\frac{dV_{eff}(x)}{dx}+\sqrt{2D(x)}\;\xi(s).  
\end{equation}
The rescaled modified potential $V_{eff}(x)$ and the modified diffusion coefficient $D(x)$ 
take the form
\begin{eqnarray} \label{VE(x)}
V_{eff}(x)&=&V(x) + \frac{1}{2}\lambda  B''(x),
 \\ 
\label{V(x)}
V(x)&=&\frac{1}{2}(x-x_e)^2 + B(x),
 \\ 
\label{D(x)}
D(x)&=&\beta^{-1}\; \left\{1-\lambda\beta [1+
B''(x)]\right\}^{-1}, 
\end{eqnarray}
where
\begin{eqnarray}
  \label{B}
 B(x)&=& \alpha \sum_{n=1}^{\infty}\frac{A_n(T_0)}{2n\pi} \left\{ p\cos(2n\pi x) \right.\nonumber\\
&+& \left. (1-p)\cos\left[2n\pi(x+1/2)\right]\right\}
\end{eqnarray}
with the rescaled temperature $T_0=T/T^*$. 
The remaining dimensionless parameters are: $x_e=\phi_e/\phi_0, \;
\alpha = LI_0/\phi_0, 
\;\lambda=\Lambda/\phi_0^2, \;1/\beta= k_BT/2E_m = k_0 T_0$, where the 
elementary magnetic flux energy $E_m=\phi_0^2/2L$ and 
$k_0= k_BT^*/2E_m$ is the ratio of two characteristic energies.  
The rescaled  zero-mean Gaussian white noise  $\xi(s)$ has the same statistical properties as thermal noise $\Gamma(t)$. 
The dimensionless quantum correction parameter 
\begin{equation}\label{lam}
\lambda = \lambda_0 \left[ \gamma + 
\Psi\left(1+   \frac{\epsilon}{T_0}\right)  \right],
\:
\lambda_0=\frac{ \hbar R}{\pi\phi_0^2}, 
\:
\epsilon = \frac{\hbar/2\pi CR}{k_BT^*}.
\end{equation}
The probability density $p(x, s)$ of the process (\ref{LL}) evolves according to the corresponding Fokker-Planck equation with natural boundary conditions. 
 The stationary probability density $P(x)$ can be obtained from the steady-state Fokker-Planck equation and  reads 
\begin{eqnarray}\label{ps}
P(x)=\lim_{s\rightarrow\infty} p(x,s) \propto D^{-1}(x) 
\exp\left[-\Phi(x)\right], 
\end{eqnarray}
where the generalized thermodynamic potential 
\begin{eqnarray}\label{phi}
\Phi(x) =\int\frac{V_{eff}'(x)}{D(x)}\;dx.
\end{eqnarray}
 Due to both the $x$-dependence of the modified diffusion coefficient 
$D(x)$ and the temperature dependence of the modified  potential 
$V_{eff}(x)$,   the stationary state (\ref{ps}) is a thermal equilibrium state, 
however,  it is not  a Gibbs state: $P_G(x) \propto \exp[-\beta V(x)]$. 

%
%
\begin{figure}[htpb]
  \begin{center}
    \includegraphics[width=0.45\textwidth]{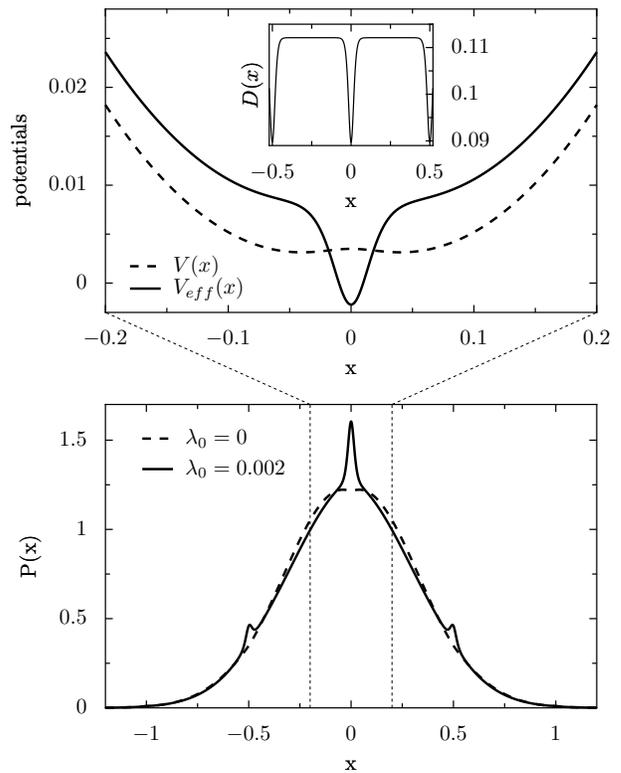}
  \end{center}
  \caption{The quantum Smoluchowski regime is compared to its 
classical counterpart.  The potential   $V(x)$ given by 
(\ref{V(x)}) and  
 the modified  potential $V_{eff}(x)$ in (\ref{VE(x)})
are shown in the upper panel. In the inset, the modified diffusion function 
$D(x)$  defined in (\ref{D(x)})  is depicted. The 
lower panel  shows  the  stationary probability density $P(x)$ in the classical Smoluchowski ($\lambda_0=0$)
and quantum Smoluchowski ($\lambda_0=0.002$) regimes. Other parameters are set as following: $x_e=0$, 
$T_0=0.2$, $\epsilon=10$, $\alpha=0.1$, $p=0.5$,  $k_0=0.5$ and $k_Fl=0.001$.}
  \label{fig1}
\end{figure}

\subsection{Quantum-renormalization of  potential and diffusion coefficient}

In Fig. 1 and 2, we present  the influence of quantum corrections on 
the shape of the potential and diffusion coefficient. 
We compare the potential $V(x)$ and  the modified quantum
potential $V_{eff}(x)$  with
each other, as well as by analyzing the modified  diffusion function
$D(x)$ (which is constant in the classical Smoluchowski domain). 
In the regime presented in Fig. 1, 
the potential $V(x)$  (dashed line) is bistable
and possesses the  barrier in contrary to $V_{eff}(x)$ (solid 
line) and  the generalized thermodynamic potential 
$\Phi(x)$ (not shown in the figure) which are  monostable and barrier-less.
The state-dependent modified diffusion function $D(x)$ possesses maxima and
minima.  The maxima and minima can be interpreted as  higher and lower effective local temperatures.  It means that quantum
fluctuations mimic a state-dependent periodic effective temperature. 
 For the escape dynamics the generalized 
thermodynamic potential $\Phi(x)$ is decisive: It contains the
combined influences of the modified  potential and the modified 
diffusion. In the regime  presented in Fig. 1,  $\Phi(x)$ has the same properties as the modified quantum potential $V_{eff}(x)$.  

The regime shown in Fig. 2 is much more interesting. 
The potential $V(x)$  (dashed line) is also bistable
and possesses the  barrier. However, the modified potential 
 $V_{eff}(x)$ (solid 
line) and $\Phi(x)$ (not shown) are  now multistable and possess many  barriers. In fact, they possess infinitely many barriers and their heights are smaller and smaller as absolute value of the flux increases. As in the previous case, the state-dependent modified diffusion function $D(x)$ possesses maxima and
minima which now are more distinct.   

Values of parameters in Figs. 1 and 2 seem to be feasible. 
A part of  values of parameters  have been evaluated from experimental data. E.g., following Mohanty \cite{mohanty}, $T^* \approx 170 mK$ and 
$T > 5mK$. Therefore the rescaled temperature  $T_0 > 0.03$.  From Ref. 2,
 we have estimated the quantum correction parameter $\lambda_0$. 
The parameters $\alpha$ and $k_0$ can be related with each other.  
The value of the parameter $\epsilon$ is unconfirmed. Fortunately, 
it enters only into the quantum correction parameter $\lambda_0$ which depends 
weakly (logarithmically) on it, cf. Eq.  (\ref{ln}). 

\rys{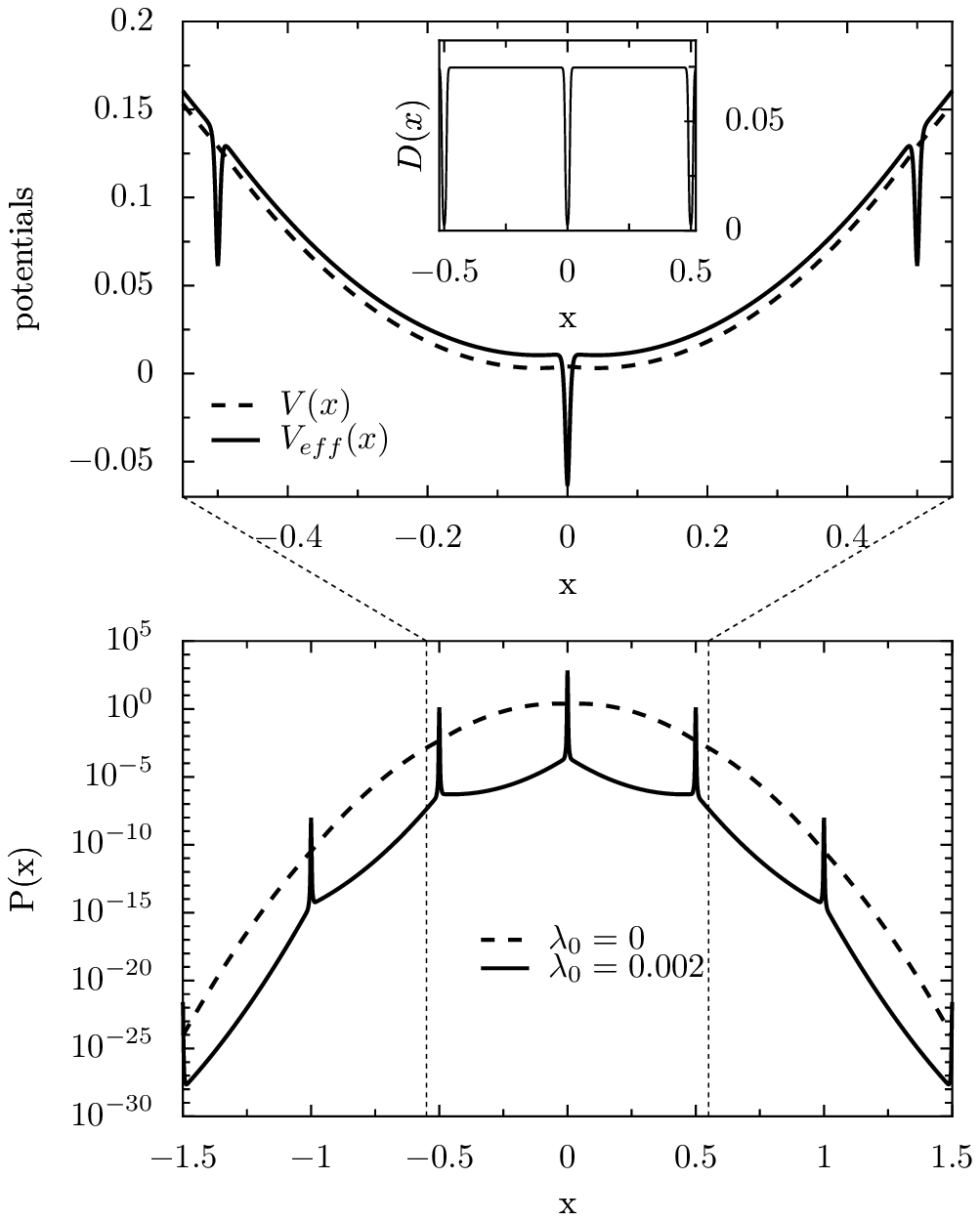}{}{The quantum Smoluchowski regime is compared to its 
classical counterpart.  The potential   $V(x)$ given by 
(\ref{V(x)}) and  
 the modified  potential $V_{eff}(x)$ in (\ref{VE(x)})
are shown in the upper panel. In the inset, the modified diffusion function 
$D(x)$  defined in (\ref{D(x)}) is depicted. The 
lower panel  shows  the  stationary probability density $P(x)$ in the classical Smoluchowski ($\lambda_0=0$)
and quantum Smoluchowski ($\lambda_0=0.002$) regimes. Other parameters are set as following: $x_e =0$, 
$T_0=0.04$, $\epsilon=10$, $\alpha=0.1$, $p=0.5$, $k_0=0.5$ and $k_Fl=0.001$.}

\subsection{Multistability}

In the following discussion we focus on the {\it self-sustaining fluxes}. 
Such fluxes, contrary to the SQUID's, has not been observed in mesoscopic rings so far. Therefore, there is a question if it may be due to additional (quantum) noise in the system. In the noiseless system, they are related to minima of the multistable generalized potential \cite{daj1}.  

In the regime where quantum corrections are negligible ($\lambda\rightarrow 0$), it is  a one-to-one correspondence between minima of the potential $V(x)$  and the maxima of the stationary probability density $P(x)$ \cite{daj1}. It is clearly not the case in the quantum  Smoluchowski regime as the modified diffusion coefficient is  flux-dependent. Nevertheless, we relate the formation of the self-sustaining currents to the appearing of the multi-peaked probability density at sufficiently low temperatures.  
As the steady state is always reflection invariant, self -sustaining fluxes are in fact  {\it  finitely-long-living}  and appear if the peaks of the steady-state probability distribution are sufficiently high. 

Let us consider two qualitatively different regimes. The first is the moderate temperature regime where the noiseless system is bistable. It is shown in Fig.1. For this case, the system, if it can be described in terms of the  'quantum Smoluchowski' equation,  is not able to accommodate self-sustaining flux  due to the destructive role of quantum fluctuations since the steady state is effectively mono-stable. 

The second regime is the  regime presented in Fig. 2, where  the onset to the multi-stable state of a noiseless system occurs. 
This regime is accessible either  by lowering temperature or using systems with larger amplitude of persistent current,  i.e. accommodating more coherent electrons.
 Here, the quantum corrections change significantly the properties of the system. Both  $V_{eff}(x)$ and $\Phi(x)$ become multi-stable what results in multistability of the steady state. The peaks are new since they do not appear at the 'classically predicted' position but rather are shifted by approximately a quarter of flux quantum $\phi_0$. There is a natural interpretation of such peaks: if they occur at $x\neq 0$  they are related to  self-sustaining  fluxes  in the system.    Their lifetimes can be estimated using the well established first-passage time method \cite{gard}.

\subsection{Lifetimes of self-sustaining flux states}

The lifetimes of  the zero and non-zero flux stationary states depend strongly 
on relation 
between the depth of the potential well of $V_{eff}(x)$ and temperature.  Therefore they can be controlled by the system parameters. It is desirable to obtain these lifetimes much longer than the characteristic time $\tau_0$, according to which time is scaled, cf. the begining of Sec. IV. 
Let us consider the regime presented in Fig. 2. The lifetime of any stationary state  $x=x_s$  can be  calculated as the mean first passage time $\tau(x_s; a, b)$ 
to leave the interval $[a, b]$ assuming that $x_s \in [a, b]$. It depends on the interval $[a, b]$ as well as on the boundary conditions (BC). We can define 
the lifetime of the state $x_s =0$  as $\tau(0; -a, a)$ with two absorbing 
BC  at $x=\pm a$, where  $a$   is a little bit greater than the 
local maximum sticked around $x= 0.0172$. Such a calculated time 
$\tau(0; -a, a) \simeq 11 \times 10^3$.   The lifetimes of the remainder states $|x_s| >  0$  
can be defined as  $\tau(x_s; a, b)$ with  one absorbing and one reflecting 
BC.  E.g. for $x_s=1/2$, one can take $a=0.488$  (which  is on the left of the local maximum sticked around $x=0.4885$) as  an absorbing BC and $b=0.52$ as a reflecting BC. Then $\tau(1/2; a,b) \simeq 4.9 \times 10^3$. Analogously, 
$\tau(1; a,b) \simeq 1.9 \times 10^3$. 
For comparison, the mean passage time from $x_s=1/2$ into $x_s=0$ is 
$\tau(1/2 \rightarrow 0) \simeq 15 \times 10^4$ and 
 from $x_s=1$ into $x_s=1/2$ is 
 $\tau(1 \rightarrow 1/2) \simeq 3 \times 10^4$. Moreover, 
 $\tau(0 \rightarrow 1/2) \simeq 8.6 \times 10^5$ and  
$\tau(1/2 \rightarrow 1) \simeq 4 \times 10^7$.  
 As a result,  the system in this regime can  effectively be treated as 
{\it tri-stable} with the reasonable level of  accuracy.

\subsection{Statistical enhancement of the magnetic flux}

The problem of the flux amplitude is more subtle. The modified  diffusion coefficient, depicted in the insets of Fig.1. and Fig.2,   is periodic with respect to  the magnetic flux $x$. If the magnetic  flux is close to half-integer, the modified  diffusion coefficient is smaller than the 'classical' Einstein one. As a result of the  interplay between this phenomenon and the shape of the modified potential one observes {\it statistical enhancement} of the magnetic flux due to quantum noise. 
This enhancement is statistical since it allows to {\it expect} an occurrence of the flux of some amplitudes with higher {\it probability} due to quantum features  of thermal equilibrium fluctuations in the quantum Smoluchowski regime. 
This enhancement is quantitative and, contrary to different approaches \cite{szwab,kraw}, this  is a purely equilibrium effect.

\rys{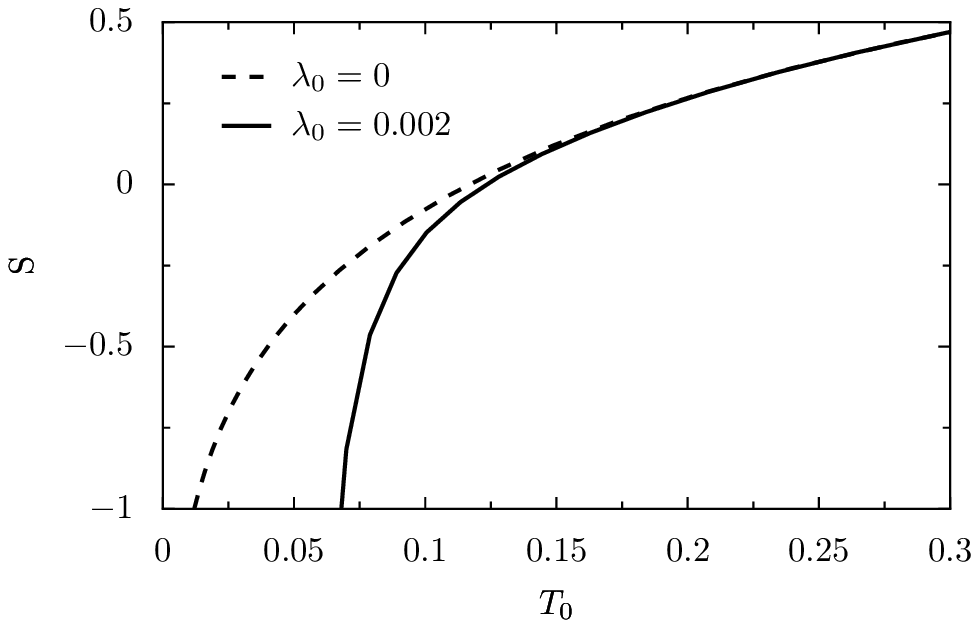}{}{'Quantum corrected' entropy (solid line) and its 
fully classical counterpart vs. temperature $T_0$.   The parameters are set as follows: $x_e =0$, 
$T_0=0.04$, $\epsilon=10$, $\alpha=0.1$, $p=0.5$, $k_0=0.5$ and $k_Fl=0.001$.}

\subsection{Relevance of peaks - entropic criterion}

There is a  question if the peaks in the multi-stable state are meaningful, i.e. if they occur in a typical experiment performed on the system. The problem can be quantified in the following equivalent way: one can ask if  the equilibrium statistics of the system is governed by ordered or quasi-ordered 'phases'.  As a measure of such a quasi-order, we exploit the celebrated Shannon entropy \cite{beck,lasota} 
\begin{eqnarray}
S[P]=-\int_{-\infty}^\infty P(x)\ln P(x) dx.
\end{eqnarray}

  Assuming a finite value of  the quantum correction parameter  
$\lambda>0$ results in decreasing of entropy,  i.e. the system becomes more ordered \cite{beck}. It is obvious that an effective order is due to increasing significance of the 'events' occurring with the high probability which are either vanishing or self-sustaining fluxes. 
We would like to stress that the entropic criterion does not characterize stability of maxima or their life-times but rather a relative frequency of their occurrence.   
The Shannon entropy plotted for two systems: with and without quantum Smoluchowski corrections is given, as a function of temperature $T_0$, in Fig.3. Working in the classical Smoluchowski regime i.e.  neglecting quantum fluctuations results in lowering an overall order in the system.     
We would like to clarify that this effect should not be interpreted as a {\it noise-induced} order. The lower entropy means simply that, contrary to the quantum Smoluchowski domain, the 'classical' regime corresponds to the disorder which is {\it over-estimated}. 

\subsection{Qutrit?}
Bistable systems are natural candidates for qubits. The celebrated examples are  Josephson-junction based devices which can be generally divided into two classes: charge and flux qubits \cite{makh}. It seems that  a qubit can also be based on 
non-superconducting materials \cite{fq}.  Because within tailored parameter regimes in the quantum Smoluchowski domain there  are symmetric peaks in the multi-stable state, such a system is a good  candidate for a {\it qutrit}. The problem of the qutrit implementation is of a central importance for quantum cryptography \cite{qtrit}.  

The following discussion is purely qualitative. We assume for simplicity that there are only three significant (in the statistical sense) peaks in  probability distribution,  as e.g. in Fig. 2. Replicating Feynman's discussion of the ammonia molecule \cite{feyn}  one can propose the 'Hamiltonian' of the system as a $3\times 3$ real symmetric matrix 
with diagonal elements  proportional to the energy of the system calculated at magnetic flux extremal value via Eq. (\ref{H}). The off-diagonal elements are proportional to the inverse of inter-peak transition times. Let us notice that in the quantum Smoluchowski regime this transitions include tunneling effects. The phenomenological modeling of quantum dynamics of the classically dissipative system may cause certain difficulties: one arrives directly at quantum dissipative system which 'conservative' component may be chosen, to some extent, arbitrary. The system under consideration can be effectively truncated to the 'qutrit'  and it is a {\it mesoscopic example} of the generic $V$-system \cite{sir}. Such a 
 system controlled by external coherent driving, i.e. equipped with an auxiliary bosonic field(s) can be naturally studied via  quantum jump approach \cite{sir}.     


\section{Conclusions}

A steady state of the magnetic flux in mesoscopic rings is both qualitatively and quantitatively different in the classical and quantum Smoluchowski regimes. Quantum effects are responsible, in dependence of parameters values,  for both destruction of bistability at moderate temperatures and formation of $n$-stability, with $n$ odd, at low temperatures. The nontrivial flux dependence of the steady state results in statistical enhancement of fluxes  of certain amplitudes. This qualitative effect is caused by  equilibrium quantum noise. 
Validity of the multi-stability has been verified via the entropic criterion. We showed that the quantum Smoluchowski regime is more ordered compared to the classical counterpart.  As the mesoscopic ring is formally identical to the zero-capacitance SQUID, it seems that the quantum Smoluchowski regime is a valid regime for  wide range of the parameters of the system and hence the effects described in the paper are of importance in  experiments performed on mesoscopic rings which are multistable systems. 

According to the {\it 'today'} common wisdom solid state devices seem promising  for implementation  of  quantum computers. Both theoretical and experimental   effort are mainly directed on superconducting  qubits. They are relatively stable with respect to decoherence and are relatively accessible. Formation of the  flux qubits in superconducting ring with a junction requires an external bias which shifts the system into the bistable state. It is not the case for the rings considered in the paper and our results  can be of importance for possible qutrit architecture based on the non-superconducting devices. Such devices, due to  their small diameters, can effectively become decoupled from  the magnetic environment \cite{fq}. This may equilibrate an absence  of the superconducting phase with its collective properties. Its is clear that capacitance, resistance, and coherent currents are the properties of the {\it whole}  non-superconducting mesoscopic rings which are thus candidates for highly integrated quantum or semi-classical circuits \cite{devor}.

\section*{Acknowledgment}
The work  supported by  the  ESF Program 
{\it Stochastic Dynamics: fundamentals and applications} and the Polish Ministry of Science and Higher Education 
under the grant N 202 131 32/3786.

\end{document}